\documentclass[prd,aps,tightline,twocolumn,nofootinbib]{revtex4}
\usepackage{graphicx}
\usepackage{amssymb}
\usepackage{amsmath}
\usepackage{color}

\begin{document}
 
\newcommand{\be}{\begin{equation}}
\newcommand{\ee}{\end{equation}}
\newcommand{\aprime}{\mathbf{a}^{\prime}}
\newcommand{\bprime}{\mathbf{b}^{\prime}}
\newcommand{\kh}{\hat{k}}
\newcommand{\Ip}{\vec{I}_+}
\newcommand{\Imi}{\vec{I}_-}
\newcommand{\bc}{\begin{cases}}
\newcommand{\ec}{\end{cases}}
\newcommand{\cD}{\mathcal{D}}
\newcommand{\xv}{\mathbf{x}}
\newcommand{\qv}{\mathbf{q}}
\newcommand{\pv}{\mathbf{p}}
\newcommand{\trec}{t_{\mathrm{rec}}}
\newcommand{\trei}{t_{\mathrm{rei}}}

\newcommand{\red}{\color{red}}
\newcommand{\cyan}{\color{cyan}}
\newcommand{\blue}{\color{blue}}
\newcommand{\magenta}{\color{magenta}}
\newcommand{\yellow}{\color{yellow}}
\newcommand{\green}{\color{green}}
\newcommand{\rem}[1]{{\bf\blue #1}}

%\begin{flushright}
%ICRR-Report-??? \\
%IPMU ??? \\
%KEK-Cosmo-???
%\end{flushright}

%\vskip 1cm

\title{
Inflation from a Supersymmetric Axion Model
}

%\vskip 1cm

\author{Masahiro Kawasaki$^{(a, b)}$, Naoya Kitajima$^{(a)}$ 
and Kazunori Nakayama$^{(c)}$}

\affiliation{%
$^a$Institute for Cosmic Ray Research,
     University of Tokyo, Kashiwa, Chiba 277-8582, Japan\\
$^b$Institute for the Physics and Mathematics of the Universe, 
     University of Tokyo, Kashiwa, Chiba 277-8568, Japan\\
$^c$KEK Theory Center, Institute of Particle and Nuclear Studies, KEK, 
     Tsukuba, Ibaraki 305-0801, Japan
}

\date{\today}

\vskip 1.0cm

\begin{abstract}
We show that a supersymmetric axion model naturally induces a hybrid inflation
with the waterfall field identified as a Peccei-Quinn scalar.
The Peccei-Quinn scale is predicted to be around $10^{15}$GeV for reproducing the large-scale
density perturbation of the Universe.
After the built-in late-time entropy-production process, the axion becomes a dark matter candidate.
Several cosmological implications are discussed.
\end{abstract}

 \maketitle

%%%%%%%%%%%%%%%%%%%%%%%%%%%%%%%%%%%%%%%%%%%%%% 
 \section{Introduction}
 %%%%%%%%%%%%%%%%%%%%%%%%%%%%%%%%%%%%%%%%%%%%%%

 The standard model in particle physics has succeeded to describe the physics below the 
 electroweak scale.
 It is not, however, a complete theory because of the theoretical problems.
 One is the strong CP problem.
 Despite the expectation of the existence of the CP-violating $\theta$-term in the Lagrangian,
 $\mathcal L = (\theta g_s^2/32\pi^2)G_{\mu\nu}^a\tilde G^{\mu\nu a}$ with $\theta \sim \mathcal O(1)$,
 experimentally it is constrained as  $\theta \lesssim 10^{-10}$.
 This is solved by the Peccei-Quinn (PQ) mechanism~\cite{Peccei:1977hh,Kim:1986ax},
 in which the axion, pseudo-Nambu-Goldstone boson associated with the spontaneous breakdown of the
 global U(1)$_{\rm PQ}$ symmetry, dynamically relaxes the $\theta$ to nearly zero.
 Another problem is the gauge hierarchy problem, which states that the 
 huge difference between the weak scale and the grand unified theory scale requires unnatural fine tuning.
 In the framework of supersymmetry (SUSY)~\cite{Martin:1997ns},
this problem does not arise.
Thus it is well motivated that we consider the SUSY axion model.

On the other hand, cosmological observations
revealed that the Universe started with inflationary expansion era 
and is now filled with unknown matter, called dark matter~\cite{Komatsu:2010fb}.
The cosmological inflation and dark matter cannot be accounted for in the framework of the standard model,
and hence they strongly indicate the physics beyond the standard model.

In this letter we point out that inflation naturally takes place in the SUSY axion model.
It takes the form of hybrid inflation where the waterfall field is identified with the PQ scalar and
the end of inflation is the PQ phase transition.
The idea was already noticed in Ref.~\cite{Copeland:1994vg}, but the discussion there was far from complete,
considering developments on the SUSY hybrid inflation model 
thereafter~\cite{Dvali:1994ms,Lazarides:1996dv,Linde:1997sj,Dvali:1997uq,Senoguz:2004vu,Jeannerot:2005mc,BasteroGil:2006cm,urRehman:2006hu,Pallis:2009pq,Nakayama:2010xf}.
(See also Refs.~\cite{BasteroGil:1997vn,Lazarides:2010di} for somewhat similar but different models.)
%A similar but different idea is found in Ref.~\cite{BasteroGil:1997vn}, 
%where no-scale type supergravity model is required for successful inflation.
%Ref.~\cite{Lazarides:2010di} considered a model which connects the inflation and PQ phase transition, 
%but each sector is separated.
Particularly, the PQ symmetry breaking scale, $f_a$, is determined as $f_a \sim 10^{15}$GeV by the condition that 
the cosmological density perturbation is in a correct magnitude.
This seems to be too large, since the axion coherent oscillation might have abundance
much larger than the dark matter.
However, we point out that due to the post inflationary dynamics of the flat direction in the scalar potential,
the axion is diluted, and it can take a role of dominant component of the dark matter.
Therefore this model provides a simultaneous solution to the hierarchy and strong CP problems,
inflation and dark matter in a simple and unified framework.

 %%%%%%%%%%%%%%%%%%%%%%%%%%%%%%%%%%%%%%%%%%%%%% 
 \section{Model and cosmological implications}
 %%%%%%%%%%%%%%%%%%%%%%%%%%%%%%%%%%%%%%%%%%%%%%

Our model is described by the following K\"ahler and superpotential, 
\begin{gather}
	K = |S|^2 + |\Psi|^2 + |\bar \Psi|^2,\\
	W = \kappa S (\Psi \bar \Psi - f_a^2) + \lambda \Psi X\bar X + kS Y\bar Y + W_0,    \label{superpot}
\end{gather}
where $S$, $\Psi$ and $\bar \Psi$ are gauge singlets,
$X (\bar X)$ and $Y (\bar Y)$ have some gauge charges, and
$\kappa$, $\lambda$ and $k$ are coupling constants, which are taken to be real and positive.
Here we keep minimal K\"ahler potentials only, and effects of non-minimal terms will be discussed later.
The constant term $W_0 (=m_{3/2}M_P^2$ where $m_{3/2}$ denotes the gravitino mass
and $M_P$ is the reduced Planck scale)
ensures that the cosmological constant is nearly zero in the present Universe.
This superpotential possesses a global U(1)$_{\rm PQ}$ symmetry,
%with charge assignments $S(0), \Psi (1), \bar \Psi(-1), X(-1/2), \bar X(-1/2)$, 
which is anomalous at the quantum level, and also has the U(1)$_R$ symmetry 
%with charge assignments $S(2), \Psi (0), \bar \Psi(0), X(1), \bar X(1)$
whose charge assignments are shown in Table.~\ref{table}.
After $\Psi$ and $\bar \Psi$ obtain vacuum expectation values (VEV),
this PQ symmetry is spontaneously broken and there appears a pseudo-Nambu-Goldstone boson,
which dynamically cancels the strong CP phase and solves the strong CP problem.

This is nothing other than the SUSY version of the hadronic (or KSVZ) axion model~\cite{Kim:1979if},
if $X$ and $\bar X$ have color charge.
In this case we can choose $Y$ and $\bar Y$ as minimal SUSY standard model
(MSSM) Higgses : $Y=H_u$ and $\bar Y = H_d$.
For a certain choice of $k$, a sizable $\mu$-term is generated 
after $S$ gets a VEV~\cite{Dvali:1997uq}, as we will see later.

It is also possible to choose $X$ and $\bar X$ to be MSSM Higgses : $X = H_u, \bar X = H_d$.
In this case, the present model describes the SUSY version of the DFSZ axion model~\cite{Dine:1981rt}.
%In the DFSZ model, standard model particles also have nonzero PQ charges.
In this case $Y$ and $\bar Y$ are additional chiral supermultiplets.
It is also allowed to introduce some additional chiral supermultiplets like heavy quarks in the KSVZ model.
In order to maintain gauge coupling unification, these additional multiplets may belong to
fundamental representations of SU(5).

One may notice that the first term in the superpotential (\ref{superpot}) 
introduced to stabilize the PQ scalar at large field value coincides with that 
used for the hybrid inflation~\cite{Copeland:1994vg,Dvali:1994ms},
after identification of $S$ with the inflaton and $\Psi (\bar \Psi)$ with the waterfall fields.
Thus we reach the interesting possibility :
the PQ sector for solving the strong CP problem naturally causes inflation.
We do not need any additional fields and interactions.
%The hybrid inflation model in supergravity was extensively studied in 
%Refs.~\cite{Senoguz:2004vu,BasteroGil:2006cm,urRehman:2006hu,Pallis:2009pq,Nakayama:2010xf}.
According to a recent analysis including the effect of constant term in the superpotential (\ref{superpot})
\cite{Nakayama:2010xf},
the correct magnitude of the density perturbation is reproduced for 
$f_a \sim 10^{15}$GeV and $\kappa \sim 10^{-3}$ if $m_{3/2}\simeq 1$~TeV.
%(Note that $k > \kappa$ is necessary, otherwise the tachyonic instability develops first toward the
%$H_u (H_d)$ direction during inflation, and leads to another unwanted minimum.)
At first sight this may seem to be disappointing, because such large PQ symmetry breaking scale
leads to axion overproduction, as is well known~\cite{Preskill:1982cy,Turner:1985si}.
In this inflationary scenario, the PQ symmetry is restored during inflation and broken after that.
Thus the phase of the axion takes random values for different patch of the Universe,
and it is not allowed to tune the initial misalignment angle to avoid the axion overproduction.

However, the situation is much better than the first thought.
This is because the late-time entropy production mechanism,
which dilutes the axion abundance to the acceptable level, is already built in the present model.
Therefore, the large PQ scale, $f_a \sim 10^{15}$GeV, is rather an appealing feature
considering that the axion can take a role of the dominant component of dark matter
after the entropy-production process.

%%%%%%%%%%%%%%%% table %%%%%%%%%%%%%%%%%%%%%%
\begin{table}[t]
  \begin{center}
    \begin{tabular}{ | c | c | c | c | c | c | c | c |}
      \hline 
       ~                               & $S$ & $\Psi$ & $\bar \Psi$ & $X$    & $\bar X$ & $Y$ & $\bar Y$ \\
       \hline \hline
        U(1)$_{\rm PQ}$  &    0   &  $+1$  &      $-1$       & $-1/2$ & $-1/2$    & $0$ & $0$ \\
      \hline
        U(1)$_R$              & $+2$ &     0   &          0          & $+1$   &  $+1$      & $0$ & $0$ \\
      \hline 
    \end{tabular}
    \caption{ 
    	Charge assignments on the field content.
           }
    \label{table}
  \end{center}
\end{table}
%%%%%%%%%%%%%%%%%%%%%%%%%%%%%%%%%%%%%%%%%%%%%% 

Now we discuss the scalar field dynamics after inflation.
The scalar potential is given by 
\begin{equation}
	V = \kappa^2 | \Psi \bar\Psi - f_a^2 |^2 + \kappa^2 |S|^2 ( |\Psi|^2 + |\bar\Psi |^2 ).
\end{equation}
Here we have taken $X=\bar X=0$ since they quickly settle at the origin due to the Hubble mass term
during inflation.
%We have also neglected the cross term coming from the constant term in the superpotential,
%because it is irrelevant in the discussion of post-inflationary dynamics.
The global minimum is located at $S=0$ and $\Psi \bar\Psi = f_a^2$.
In other words, there is a flat direction along which the scalar fields do not feel the potential,
ensured by the U(1)$_{\rm PQ}$ symmetry extended to a complex U(1) due to the holomorphy.
%The existence of the flat direction is related to the holomorphy of the superpotential :
%the real U(1)$_{\rm PQ}$ symmetry is extended to the complex U(1)$_{\rm PQ}$ symmetry
%due to the holomorphy of the superpotential and it is nothing but the 
%invariance under a conformal transformation.
The SUSY breaking effect lifts up the flat direction, saxion, and gives a mass of order $m_{3/2}$,
\begin{equation}
	V_{\rm SB} = c_1 m_{3/2}^2 |\Psi|^2 + c_2  m_{3/2}^2 |\bar \Psi|^2,   \label{VSB}
\end{equation}
where $c_1$ and $c_2$ are $\mathcal O (1)$ constants.
This stabilizes the flat direction at $|\Psi| \simeq |\bar\Psi| \simeq f_a$.
We denote deviation from this minimum along the flat direction as $\sigma$, and call it as saxion.
The $\Psi$ field also receives a finite-temperature effective potential, $V_T \simeq \alpha_s^2 T^4 \log \Psi$,
where $\alpha_s$ is the QCD gauge coupling constant,
coming from two-loop effects even if heavy quarks are decoupled from thermal bath~\cite{Anisimov:2000wx}.

After inflation ends, the inflaton $S$ and waterfall fields $\Psi (\bar \Psi)$ oscillates around the minimum,
$S = 0, |\Psi| =|\bar\Psi| = f_a$, noting that the flat direction at this stage obtains a mass of $\kappa |S|$.
The scalar degrees perpendicular to this direction, 
which fully mixes with $S$, decays much earlier than the saxion 
since they have masses of $m_S \sim \kappa f_a$.
The decay is induced by the third term in (\ref{superpot}), and the reheating temperature is 
around $T_{\rm R} \sim 10^{11}$GeV for $m_S \sim 10^{12}$GeV and $k\sim \kappa$.
After that, the thermal logarithmic comes to dominate and drives the saxion to
$|\bar \Psi| \sim \alpha_s M_P$ where the effective thermal mass becomes equal to the Hubble parameter,
and the saxion stops there until the thermal effect becomes irrelevant.
When the Hubble parameter decreases to $\sim m_{3/2}$, 
the mass term dominates over the thermal correction, and the saxion begins to oscillate around the
minimum, $|\Psi|\sim |\bar \Psi| \sim f_a$, with an initial amplitude of $\sigma_i \sim \alpha_s M_P$.
%After that, thermal logarithmic potential comes to dominate the saxion dynamics and 
%oscillation along the flat direction begins, with amplitude $\sim \sqrt{\alpha_s}T_{\rm R}^2/m_{3/2}$.
%Then the amplitude gradually decreases, 
%and when the effective thermal mass becomes equal to $\sim m_{3/2}$,
%the saxion oscillates around the true minimum with an amplitude of $\sigma_i$.
%An exact expression of $\sigma_i$ is a bit complicated, 
%but numerically it lies in the range $\sim 10^{16}$-$10^{17}$GeV, not far from $f_a$.
%Thus we parametrize it by $f_a$ for simplicity.
The abundance of the saxion coherent oscillation, 
in terms of the energy density to entropy ratio, is then given by\footnote{
	Here we have assumed that the $S$ decays before the saxion begins to oscillate.
	Otherwise, the presence of the $\kappa |S|$ mass term for the saxion makes the
	saxion oscillation amplitude exponentially suppressed~\cite{Linde:1996cx}.
}
%%
%\begin{equation}
%	\frac{\rho_\sigma}{s}
%	\simeq 2\times 10^{4}{\rm GeV} \xi \left( \frac{m_{\sigma}}{1{\rm TeV}} \right)^{1/2}
%	\left( \frac{f_a}{10^{15}{\rm GeV}} \right)^2
%	\left( \frac{\sigma_i}{f_a} \right)^2,
%\end{equation}
%%
%%
\begin{equation}
\begin{split}
	\frac{\rho_\sigma}{s}
	&= \left( \frac{90}{\pi^2 g_*} \right)^{1/4} \frac{\sqrt{m_\sigma M_P}}{8}\frac{\sigma_i^2}{M_P^2} \\
	&\simeq 1\times 10^8{\rm GeV} \left( \frac{m_{\sigma}}{1{\rm TeV}} \right)^{1/2}
	\left( \frac{\sigma_i}{\alpha_s M_P} \right)^2,
\end{split}
\end{equation}
where %$\xi = (\alpha_s M_P/f_a)^{3/2} \sim 10^3$, and 
$m_\sigma (\sim m_{3/2})$ is the saxion mass.
This comes to dominate the Universe well before the QCD phase transition.
In the case of KSVZ model, the saxion decays into gluons with the rate\footnote{
	Decay into two axions, $\sigma \to 2a$, must be suppressed for successful reheating 
	in the KSVZ model. This requires $c_1\simeq c_2$ in Eq.~(\ref{VSB})
	\cite{Chun:1995hc,Kawasaki:2007mk}.
	%Taking account of the cosmological bound on the additional radiation energy,
	%the initial amplitude must be suppressed as $\sigma_i \lesssim 10^{-2} f_a$.
}
\begin{equation}
	\Gamma_{\sigma \to gg} = \frac{\alpha_s^2}{32\pi^3}\frac{m_\sigma^3}{f_a^2}.
\end{equation}
Then the Universe is reheated again by the saxion decay. 
The temperature after the saxion decay is estimated as
\begin{equation}
	T_\sigma \sim 3~{\rm MeV}\left( \frac{m_\sigma}{10{\rm TeV}} \right)^{3/2}
	\left( \frac{10^{15}{\rm GeV}}{f_a} \right),
\end{equation}
and hence is compatible with the lower bound on the reheating temperature~\cite{Kawasaki:1999na}
for $m_\sigma \gtrsim 10$TeV.
Notice that the saxion also decays into a SUSY particle pair and then produces 
the lightest SUSY particles (LSP) nonthermally,
which easily exceed the dark matter abundance.
Thus we need to introduce small R-parity violation in order for the LSP to decay well before BBN begins,
or to assume SUSY particles are heavy enough not to be produced by the saxion decay.
In the case of DFSZ model, the saxion decays into Higgs pair or fermion pairs.
For example, the decay width into the lightest Higgs boson pair is
\begin{equation}
	\Gamma_{\sigma \to hh} = \frac{1}{8\pi}\frac{m_\sigma^3}{f_a^2} \left( \frac{\mu}{m_\sigma} \right)^4,
\end{equation}
where $\mu = \lambda \langle \Psi \rangle$ gives the higgsino mass, and hence we obtain
\begin{equation}
	T_\sigma \sim 5~{\rm MeV}\left( \frac{m_\sigma}{1{\rm TeV}} \right)^{3/2}
	\left( \frac{10^{15}{\rm GeV}}{f_a} \right) \left( \frac{\mu}{m_\sigma} \right)^2.
\end{equation}
Thus in this case we need $m_\sigma \sim 1$~TeV and 
decay into a SUSY particle pair can naturally be forbidden.

%%%%%%%%%%%%%%%
%\begin{figure}[htbp]
%\begin{center}
%\includegraphics[width=0.8\linewidth]{oscillation.eps}
%\caption{ The dynamics of PQ scalar, $\Psi$ and $\bar \Psi$ as a function of time $t$. 
%		Here we have taken $\kappa = 0.01$, $f_a=10^{15}$GeV, $c_1=2$, $c_2=3$ and
%		$m_{3/2}=10^{11}$GeV. The gravitino mass is taken to be this large value for the purpose of %numerical calculation. }
%\label{fig:osc}
%\end{center}
%\end{figure}
%%%%%%%%%%%%%%%

Now let us discuss the abundance of the axion, gravitino and axino after the dilution by the saxion decay.
The axion abundance, in terms of the density parameter, 
after the dilution is estimated as~\cite{Steinhardt:1983ia}
\begin{equation}
	\Omega_a h^2 \simeq 5\times 10^{-2} \left( \frac{T_\sigma}{1{\rm MeV}} \right)
	\left( \frac{f_a}{10^{15}{\rm GeV}} \right)^2,
\end{equation}
hence it is consistent with the WMAP observation of the dark matter abundance.
This is appealing, since the PQ scale $f_a \sim 10^{15}$GeV is required for generating
the density perturbation of the Universe, while this large PQ scale leads to the efficient late-time
entropy production, making the axion plausible candidate of dark matter.
Note that the axion does not have an isocurvature perturbation in this model,
since the PQ symmetry is restored during inflation.

As for the gravitino, 
%both nonthermal production from the inflaton decay~\cite{Kawasaki:2006gs}, 
%and thermal production~\cite{Bolz:2000fu}, are diluted sufficiently.
they are produced both thermally~\cite{Bolz:2000fu} and nonthermally~\cite{Kawasaki:2006gs}
from the inflaton decay, but diluted sufficiently.
The thermally produced gravitino abundance, in terms of the number to entropy ratio, is estimated as
%%
%\begin{equation}
%	Y_{3/2} \simeq \frac{1\times 10^{-22}}{ (m_\sigma/{\rm TeV})^{1/2} }
%	\frac{T_{\rm R}}{10^{11}{\rm GeV}} 
%	\frac{T_\sigma}{1{\rm MeV}} 
%	\left( \frac{\alpha_s M_P}{\sigma_i} \right)^2.
%\end{equation}
%%
%%
\begin{equation}
\begin{split}
	Y_{3/2} \simeq & 1\times 10^{-22}
	\left( \frac{1{\rm TeV}}{m_\sigma} \right)^{1/2} \\
	& \times \left (\frac{T_{\rm R}}{10^{11}{\rm GeV}} \right )
	\left ( \frac{T_\sigma}{1{\rm MeV}}  \right )
	\left( \frac{\alpha_s M_P}{\sigma_i} \right)^2.
\end{split}
\end{equation}
This satisfies the bound on the unstable gravitino abundance from BBN 
$m_{3/2}Y_{3/2}\lesssim 10^{-13}$-$10^{-9}$GeV 
for $m_{3/2}\sim 1$-10TeV~\cite{Kawasaki:2004yh,Jedamzik:2006xz}
for an unstable gravitino.
 
The axino, which is the fermionic superpartner of the axion, might also have
significant effects on cosmology~\cite{Rajagopal:1990yx,Goto:1991gq}.
The axino abundance from thermal production~\cite{Covi:2001nw}, after the dilution, is given by
%%
%\begin{equation}
%	Y_{\tilde a} \simeq \frac{1\times 10^{-19}}{ (m_\sigma / {\rm TeV})^{1/2}}
%	\frac{T_{\rm R}}{10^{11}{\rm GeV}}
%	\frac{T_\sigma}{1{\rm MeV}}
%	\left( \frac{10^{15}{\rm GeV}}{f_a} \right)^2
%	\left( \frac{\alpha_s M_P }{\sigma_i} \right)^2.   \label{axino}
%\end{equation}
%%
%%
\begin{equation}
\begin{split}
	Y_{\tilde a} \simeq & 1\times 10^{-19}
	\left( \frac{1{\rm TeV}}{m_\sigma} \right)^{1/2}
	 \left( \frac{T_{\rm R}}{10^{11}{\rm GeV}} \right) \\
	&\times \left( \frac{T_\sigma}{1{\rm MeV}} \right)
	\left( \frac{10^{15}{\rm GeV}}{f_a} \right)^2
	\left( \frac{\alpha_s M_P }{\sigma_i} \right)^2.   \label{axino}
\end{split}
\end{equation}
In the present model, the axino mass is generated once the $A$-term potential is included :
$V_A = A \kappa S f_a^2 +{\rm h.c.}$ with $A \sim m_{3/2}$.
Then $S$ has a VEV of $\sim A/\kappa$, and it gives an axino mass of 
$m_{\tilde a}=\kappa \langle S \rangle \sim A$. Thus the axino mass is comparable to the gravitino.
If the axino is not the LSP, it has a similar lifetime to the saxion in the KSVZ model, 
and it decays before BBN. The constraint is given as $Y_{\tilde a} \lesssim 10^{-12}$
so as not to produce too much LSPs.
If the axino is the LSP, the bound reads $m_{\tilde a}Y_{\tilde a} \lesssim 4\times 10^{-10}$GeV.
In both cases, the constraint is satisfied as is seen in Eq.~(\ref{axino}).
%In the KSVZ model, the axino has a similar lifetime to the saxion and decays slightly before BBN.
%Since the R-parity should be violated, LSPs produced by the axino decay soon disappear.
%In the DFSZ model, the axino may be the LSP, and the bound reads 
%$m_{\tilde a}Y_{\tilde a} \lesssim 4\times 10^{-10}$GeV. This is satisfied for natural parameter choices.

%Then the bound reads $m_{\tilde a}Y_{\tilde a} \lesssim 10^{-10}$GeV
%~\cite{Kawasaki:2004yh,Jedamzik:2006xz}.
%This is also satisfied for reasonable parameter choices.
%If the axino is the LSP, the bound is given as $m_{\tilde a}Y_{\tilde a} \lesssim 4\times 10^{-10}$GeV.

%Finally, the LSP, which may be the lightest neutralino,
%is produced at the thermal freeezeout at $T \sim \mathcal O(10)$
%GeV and also by the decay of gravitino and axino.
%The former conribution is diluted by the saxion decay, and the latter contribution is also small
%because the parent gravitino and axino are diluted already as shown above. 
%Thus the LSP abundance is expected to be smaller than the dark matter abundance.

%%%%%%%%%%%%%%%%%%%%%%%%%%%%%%%%%%%%%%%%%%%%%% 
 \section{Discussion} \label{disc}
 %%%%%%%%%%%%%%%%%%%%%%%%%%%%%%%%%%%%%%%%%%%%%%

Here we briefly discuss several remaining issues.

{\it Coleman-Weinberg correction :}
Since the waterfall fields couple to $X (\bar X)$,
there arises a Coleman-Weinberg (CW) effective potential for the $\Psi$ direction~\cite{Coleman:1973jx}.
We have explicitly checked that this does not modify the (post-)inflationary scalar dynamics at all.
Also the presence of $Y (\bar Y)$ affects the potential of the inflaton ($S$) through the CW correction.
As long as $k$ is not much larger than $\kappa$, the inflaton dynamics is not much affected.
Since we need $k>\kappa$ for relaxing to desired vacuum~\cite{Dvali:1997uq}.
In this letter we have assumed $k \sim \kappa$.

{\it Stability of heavy quarks :}
In the hadronic axion model, the heavy quarks are stable unless some additional
operators are introduced. Thus the existence of too much heavy quarks may be problematic.
In our model, however, they are not produced in the early Universe efficiently,
since the reheating temperature is lower than the heavy quark mass scale.
In the DFSZ model, $Y (\bar Y)$ have weak scale masses and may also be stable, 
and are once thermalized after the reheating.
But its relic abundance is significantly reduced by the saxion decay and no significant cosmological effects arise.

{\it Non minimal K\"ahler potentials :}
In the hybrid inflation model in supergravity,
the scalar spectral index $n_s$ takes a value from $0.98$-$1.00$
for the range of $f_a\sim 10^{15}$GeV-$10^{16}$GeV in the minimal K\"ahler potential
\cite{BasteroGil:2006cm,Nakayama:2010xf}.
It is possible to make the spectral index more red tilted and fall into the best fit range
($n_s =0.963\pm0.012$ with 68\% C.L.~\cite{Komatsu:2010fb}),
by introducing a non-minimal K\"ahler potential $K = k_S |S|^4/M_P^2$
and choosing its coefficient as $k_S \sim 0.01$-0.02.
%On the other hand, the tensor-to-scalar ratio, $r$, is as small as $r \sim 10^{-11}$
%and the detection of primordial gravitational wave background is hopeless.
For example, $n_s=0.96$ $(0.95)$ is obtained for $f_a=10^{15}$GeV
$(4\times 10^{15}$GeV) for $k_S=0.01$
~\cite{BasteroGil:2006cm,Nakayama:2010xf}.
We can also introduce non-minimal K\"ahler potentials, as
$K = (k_1|S|^2|\Psi|^2+k_2|S|^2|\bar\Psi|^2)/M_P^2$ with $k_1\sim k_2\sim \mathcal O(1)$.
These terms give an additional Hubble mass to the saxion.
But this does not modify the saxion dynamics at all, since the saxion mass is dominated by the 
$\kappa |S|$ term during the $S$ oscillation, and after $S$ decays these terms become irrelevant.

%Depending on the sign of $k_1$ and $k_2$, the Hubble mass can be either positive or negative. 
%In the case of positive Hubble mass, the saxion dynamics is not much affected.
%It only changes the saxion initial amplitude within an $\mathcal O(1)$ factor.
%In the case of negative Hubble mass, however, the saxion rolls down the flat direction to the Planck scale
%after the $S$ decays.
%Thus the saxion initial amplitude is expected to be of order of $M_P$, not $f_a$.
%This releases huge amount of entropy and gravitinos are diluted more efficiently~\cite{Kawasaki:2008jc}.

{\it Domain wall problem :}
After inflation U(1)$_{\rm PQ}$ is broken and cosmic strings are formed.
After the QCD phase transition, domain walls appear which is bounded by the strings.
Due to the tension of domain walls, strings (and walls) shrink and finally disappear
if the color anomaly number is one, as in the KSVZ model.
This is not the case for the DFSZ model, hence we need to introduce several heavy quarks 
in order to make the color anomaly number one.

{\it Axions emitted from strings :}
The axionic strings continuously emit axions with momentum of order of the Hubble scale.
After the QCD phase transition it obtains a mass, and the energy density stored in the axions
emitted by the strings is comparable to that from the coherent oscillation~\cite{Yamaguchi:1998gx}.
Thus these axions may contribute to some non-negligible fraction of the dark matter.

%{\it Thermal effects :}
%The PQ scalar, $\Psi$, may receive a thermal correction to the potential after inflation
%coming from the two-loop effect,
%as $V_T \sim \alpha_s^2T^4\log \Psi$~\cite{Anisimov:2000wx}, even if heavy quarks are decoupled,
%and this term is comparable to the zero-temperature potential for the saxion.
%This tends to make the saxion oscillation epoch earlier, with larger initial amplitude.
%Although the estimation of the saxion abundance receives a slight modification, the final qualitative results 
%are not affected.

{\it Baryon asymmetry :}
Because of the inevitable late-time entropy production process, 
any preexisting baryon asymmetry is diluted.
One possibility to generate the baryon asymmetry which survives the huge dilution is to rely on the
Affleck-Dine (AD) mechanism~\cite{Affleck:1984fy}.
Such a scenario was studied in detail in Ref.~\cite{Kawasaki:2007yy}.
The baryon-to-entropy ratio is expressed as
\begin{equation}
	\frac{n_B}{s} \simeq \frac{\delta_{\rm CP}m_{3/2}|\psi_{\rm os}|^2}{m_\sigma^2 \sigma_i^2 \xi}
	\frac{3T_\sigma}{4},
\end{equation}
where $\delta_{\rm CP}\sim \mathcal O(1)$ is the effective CP angle,
$\psi_{\rm os}$ denotes the AD field amplitude at the onset of its oscillation,
and $\xi$ is a suppression factor due to the mismatch of the oscillation epochs of the AD field and the saxion.
As an example, the AD field lifted by the non-renomalizable superpotential 
$W_{\rm AD} = \psi^n/nM^{n-3}$ with a cutoff scale $M$ corresponds to 
$\psi_{\rm os} =(H_{\rm os} M^{n-3})^{1/(n-2)}$,
where $H_{\rm os}$ is the Hubble parameter at the onset of AD fields oscillation.
In the present parameter, the AD fields begins to oscillate around $T=T_{\rm R}$,
and in this case we have $\xi \sim (\sqrt{m_\sigma M_P}/T_{\rm R} )^{3}$.
Substituting typical values of $T_{\rm R}\sim 10^{11}$GeV,
$\sigma_i\sim \alpha_s M_P$, $M\sim M_P$ and $n=6$
yields the correct observed value, $n_B/s \sim 10^{-10}$.
Note that Q-balls formed through the AD mechanism decay well before BBN and never dominates the Universe
because of the presence of the saxion.
It should also be noted that the baryonic isocurvature fluctuation generated by the AD mechanism 
is of the order of 
$\sim (H_{\rm inf}/M)^{3/4}$, where $H_{\rm inf}\sim \kappa M^2/M_P$ is the Hubble scale during inflation,
and is safely ignored.

To summarize, we have shown that a simple SUSY axion model naturally causes hybrid inflation.
The PQ symmetry breaking scale is fixed to be around $10^{15}$GeV 
so that the correct magnitude of cosmological density perturbation is reproduced.
Taking into account the post inflationary dynamics of the scalar fields contained in the model,
the late-time entropy production necessarily occurs and the axion coherent oscillation can be the dark matter.
This provides a solution to the strong CP and gauge hierarchy problems
and simultaneously explains a cosmic inflation and the presence of dark matter.

%%%%%%%%%%%%%%%%%%%%%%%%%%%%%%%%%%%%%%%%%%%%
\begin{acknowledgments}
%%%%%%%%%%%%%%%%%%%%%%%%%%%%%%%%%%%%%%%%%%%%

KN would like to thank Fuminobu Takahashi for useful discussion.
This work is supported by Grant-in-Aid for Scientific research from the Ministry of Education,
Science, Sports, and Culture (MEXT), Japan, No.14102004 (M.K.) 
and No. 21111006 (M.K. and K.N.)
and also by World Premier International
Research Center Initiative (WPI Initiative), MEXT, Japan.

 %%%%%%%%%%%%%%%%%%%%%%%%%%%%%%%%%%%%%%%%%%%%
\end{acknowledgments}
%%%%%%%%%%%%%%%%%%%%%%%%%%%%%%%%%%%%%%%%%%%%

%%%%%%%%%%%%%%%%%%%%%%%%%%%%%%%%%%%%%%%%%%%%
   
%%%%%%%%%%%%%%%%%%%%%%%%%%%%%%%%%%%%%%%%%%%%%
 
 \end{document}